# PC2IM: An Efficient In-Memory Computing Accelerator for 3D Point Cloud

Dengfeng Wang, Shunqin Cai, and Yanan sun

*Abstract*—**3D point cloud neural networks have significantly enhanced the perceptual capabilities of resource-limited mobile intelligent systems. However, despite the transformative impact, the point cloud algorithm suffers from substantial memory access during data preprocessing and imposes a burdensome workload on feature computing, resulting in high energy consumption and latency. In this paper, an efficient SRAM-based computing-in-memory (SRAM-CIM) accelerator (PC2IM), is proposed to alleviate memory access bottlenecks in point-based 3D point cloud networks. A data preprocessing module driven by the customized CIM engines is proposed and incorporated into a memory-efficient data flow. Specifically, an approximate distance SRAM-CIM (APD-CIM) is introduced to eliminate the repetitive on-chip memory access for point clouds that are spatially partitioned by the median and reduce the volume of temporary distance data. Building on the APD-CIM, a two-level Ping-Pong-MAX Content Addressable Memory (Ping-Pong-MAX CAM) is introduced to adaptively update temporary distances and perform in-situ search for the maximum, further reducing memory access. Additionally, an efficient CIM-based feature computing engine, named split-concatenate SRAM-CIM, is presented to minimize computation latency in multi-layer perceptron with high-precision input, while maintaining high area and energy efficiency. Experiment results show that the proposed PC2IM demonstrates 1.5× speedup and 2.7× enhanced energy efficiency compared to state-of-the-art point cloud accelerator. Moreover, PC2IM achieves 3.5× speedup and 1518.9× enhanced energy efficiency compared to GPU implementations.**

## I. Introduction

Point Clouds (PCs) acquired through LiDAR or depth sensors encapsulate the intricacies of real-world scenes with unparalleled detail. With this nuanced data representation, new frontiers are opened for applications demanding energy-efficient real-time processing, ranging from Autonomous Driving (AD) to Augmented Reality (AR) and environmental monitoring.

Point Cloud Networks (PCNs) have become pivotal to perceive and interpret the three-dimensional (3D) information of PCs [1]-[2]. To learn features from the inherently sparse and unstructured PCs, PCNs need to structure PCs with mapping operations (data preprocessing) by partitioning raw data into small point sets, then computing features with Multi-Layer Perceptrons (MLPs). Each point sets consist of a centroid and its corresponding neighbors. The centroid is typically derived using down-sampling techniques like Farthest Point Sampling (FPS) to preserve the original spatial characteristics, while neighbor points are obtained through methods like k-Nearest-Neighbors search (kNN) and ball query. However, FPS poses challenges, requiring iterative traversal of large-scale raw PCs for global down-sampling. This results in significant power and latency issues due to the frequent irregular off-chip memory transactions [13], with FPS consuming up to 70% of the total PCN runtime in large-scale PCs [3]. Additionally, MLPs in PCNs demand high-precision Multiply-and-Accumulate operations (MACs), typically exceeding 8 bits [3]-[10][13], contributing significantly to energy consumption and latency across the entire network. As a consequence, the main stream PCNs can only run at about 10 frame per second (fps) [4] with low energy efficiency [9] on a generic computing platform. These challenges underscore the difficulty in achieving energy-efficient, real-time PCN accelerations within resource-limited edge intelligent systems.

Recent advancements in PCN accelerators have effectively tackled the challenges mentioned through innovative algorithm-architecture co-design strategies [5]-[12]. MARS [5] partitions large PCs into small grids with grid grouping to filter out unnecessary off-chip data access. PNNPU [6] uses hierarchical block-wise FPS for efficient point processing. Point-wise feature reuse strategy is proposed in [7]-[9] to reduce redundant feature computations. TiPU [10] introduces a spatial-locality-aware sampling method to sample points in small fixed-shaped local tile without global PC access and adopts efficient bit-serial near-memory MAC units for MLPs. Morton-Code-based spatial partitioning (SP) methods are used to perform two-level mixed-precision quantization [11] and fused sampling and grouping [12] on PCNs. However, prior works primarily concentrate on mitigating the substantial off-chip global PC accesses through SP [5][6][10]-[12]. A new challenge arises with on-chip data transfer, which becomes the dominant factor in overall memory transactions within SP based PCNs. Furthermore, the incorporation of high-precision MLPs poses challenges for conventional bit-serial near-memory computing schemes, resulting in extended computing cycles and linearly scaled energy consumption as the length of the input sequence increases.

To address these challenges above, we propose PC2IM, an efficient SRAM-CIM accelerator to alleviate both off-chip and on-chip memory access issues in point-based 3D point cloud networks. The contributions of this work are listed as follows:

(1) A CIM-based data preprocessing module incorporated in a memory-efficient data flow is proposed for point-based PCNs, which is driven by the proposed approximate distance SRAM-CIM (APD-CIM) and a two-level Ping-Pong-MAX content-addressable memory (Ping-Pong-MAX CAM). APD-CIM is used for approximate distance based sampling and lattice query on the local PC partitioned by the median, while the two-level Ping-Pong-MAX CAM is used to efficiently perform in-situ search for the maximum with reduced temporary data access. The proposed CIM-based data preprocessing module alleviates the on-chip data transfer, significantly enhancing the energy efficiency with 73.4% reduced energy consumptions in the data preprocessing stage of PCNs compared to state-of-the-art PCN accelerator.

(2) An efficient feature computing engine, named split-concatenate SRAM-CIM (SC-CIM), is proposed to significantly reduce computing latency of conventional digital SRAM-CIM in high-precision MACs while maintaining high area and energy efficiency. Block-wise and bit-wise split schemes are employed for weights and inputs, respectively, to efficiently conduct fused multiply-and-add operations, while the partial sums are accumulated by the sparse-dense adder tree with low area overhead. The proposed SC-CIM achieves 2.0× to 9.9× enhanced Figure of Metric (FoM) compared with conventional digital SRAM-CIM, considering different storage-compute ratios.

(3) An efficient SRAM-CIM accelerator named PC2IM is proposed for PCN accelerations with enhanced performance and energy efficiency. Experiment results show that the proposed PC2IM achieves 1.5× speedups and 2.7× enhanced energy efficiency in various workloads compared to state-of-the-art PCN accelerator. Compared with GPU implementation, PC2IM achieves 3.5× speedups and 1518.9× enhanced energy efficiency on large-scale PC dataset.

## II. Background an Motivations

### A. Point Cloud Networks

Point Cloud Networks (PCNs) refer to a class of neural network architectures designed to handle point cloud data, which is a set of data points in a 3D coordinate system with optional RGB-D or intensity information [1]. PCNs can be broadly categorized into voxel-based [2] and point-based networks [1]. Voxel-based PCNs discretize point clouds into a regular grid of voxels and may not capture intricate features inherent in the original point cloud data, potentially resulting in the loss of fine-grained details. On the other hand, point-based PCNs operate directly on unstructured point cloud data without voxelization or intermediate representations. Therefore, point-based PCNs offer a promising solution for learning features from 3D point clouds. The typical structure of a point-based network is illustrated in Fig. 1 [1]. The fundamental module of a point-based PCN can be abstracted into two main components: data preprocessing and feature computing.

***Data preprocessing.*** In point-based PCNs, unstructured raw point cloud needs to be structured into small point sets for feature learning. This is done in data preprocessing stage of point set abstraction layer, where farthest point sampling and ball query/kNN are applied on raw PC. Farthest point sampling is an algorithm employed for down sampling a subset of points from a larger set based on their spatial distribution. In FPS, distance $\boldsymbol{D_s}$ of points outside the sampling set $\boldsymbol{S}$ is calculated to sample a centroid point with maximal $\boldsymbol{D_s}$ into $\boldsymbol{S}$ from raw PCs, where $\boldsymbol{D_s}$ is defined as the minimal distance of raw points to $\boldsymbol{S}$. Generally, temporary distance memory will be used to store $\boldsymbol{D_s}$, and updated after each sampling to reduce redundant distance calculations. Note that, for each sampling, the whole raw PCs need to be traversed to calculate the maximal $\boldsymbol{D_s}$, which can be very time-consuming in large-scale PCs. Followed with FPS, ball query identifies the nearest neighbor points around centroid point within a distance threshold. The centroid and its neighbors are then aggregated and send to the feature computing stages. For segmentation and detection tasks requiring characteristics of individual points, points with high-dimensional features undergo up sampling. This involves kNN search and weighting up features of neighbors in the data preprocessing stage of the point feature propagation layer.

***Feature computing.*** In feature computing stage, MLPs are applied independently to each point set, allowing the network to learn point-wise features and capture local details. The output of the MLP provides an enhanced representation of the input features for each point. Max pooling is applied after the MLPs and is used to aggregate information across a local neighborhood of points. It helps capture local structures and identify salient features within a specified region of the point cloud.

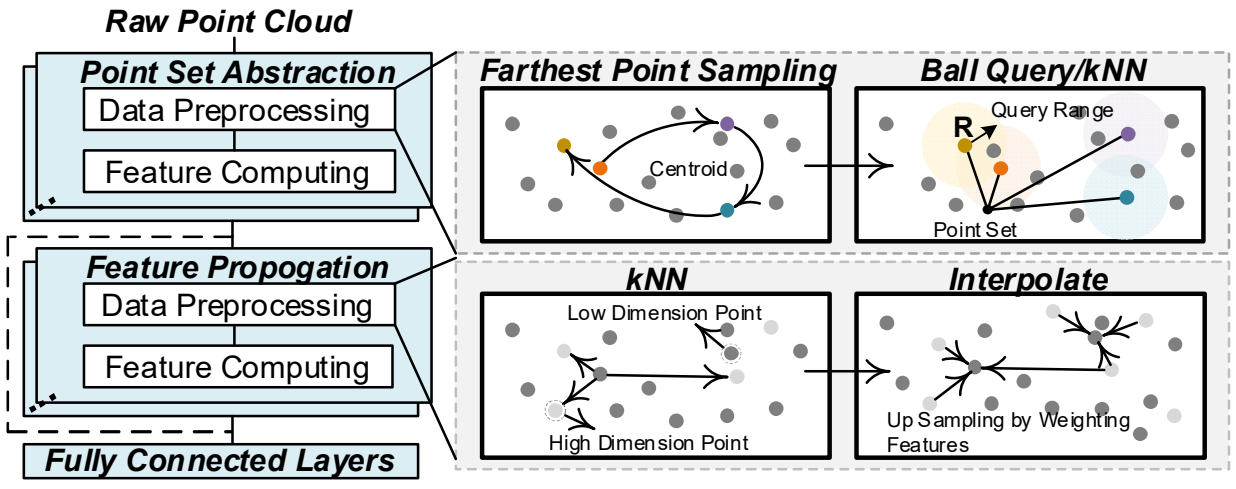

Fig. 1. The typical network structure of point-based PCN [1].

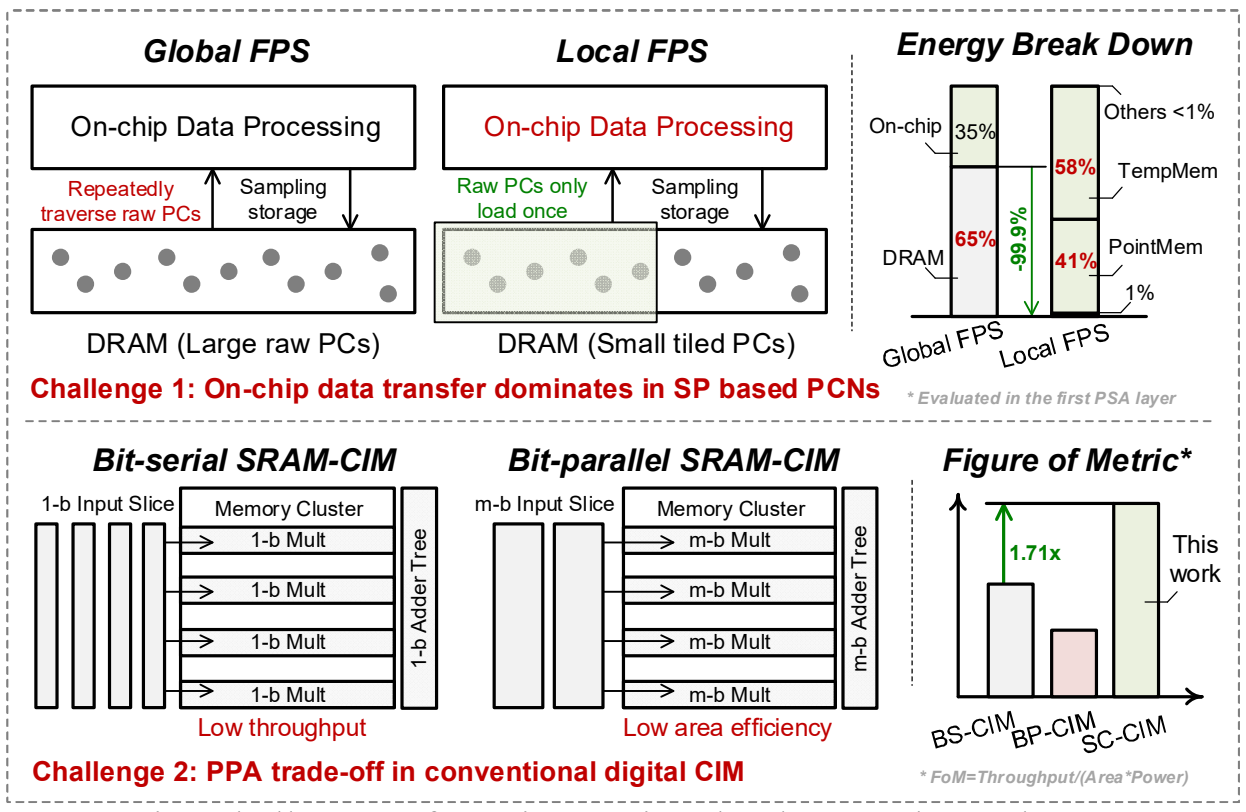

Fig. 2. Design challenges of previous point cloud network accelerator.

### B. Bottleneck and Motivations

Traditional global FPS in PCNs necessitates iterative access to global, large-scale point clouds for down sampling points, resulting in substantial off-chip memory transactions. To mitigate off-chip memory access, TiPU [10] employs a spatial-aware partitioning method, allowing for point sampling within small local tiles that fit on-chip. Additionally, TiPU [10] utilizes energy-efficient bit-serial near-memory computing (BS-SRAM-CIM) with delayed aggregation to expedite intensive MAC operations while reducing intermediate feature sizes. Despite these advancements, challenges arise with local FPS and BS-SRAM-CIM for efficient PCN acceleration, as illustrated in Fig. 2.

***Challenge I: On-chip data transfer dominates in spatial partitioning based PCNs.*** Given that the tiled PC is compact enough to be accommodated on-chip for local FPS and neighbor search, the large-scale raw PCs only need to be loaded once from DRAM to on-chip memory, resulting in a 99.9% reduction in DRAM access in SP based PCNs. However, to sample the local PC on-chip, the entire set of tiled points must be repeatedly read from on-chip memory for distance calculations. Additionally, in sampling phase, the minimal distance list requires consistently updating, incurring substantial energy consumption. Therefore, in local FPS, on-chip memory access becomes the dominant factor, accounting for 99% of overall memory access, with 41% related to on-chip point access and 58% to temporary distance update overhead.

***Challenge II: PPA trade-off in conventional digital SRAM-CIM.*** High-precision MLPs face a bottleneck in conventional BS-SRAM-CIM. BS-SRAM-CIM efficiently sends one-bit input per cycle for multiplication with 1-bit multiplier, ensuring high area efficiency. However, its performance degrades significantly when dealing with high-precision input.

Additionally, the computing energy scales linearly with input length. While processing multi-bit input in one cycle is possible, it introduces substantial area overhead. This is attributed to the need for a multi-bit multiplier for each memory cluster, and the bit-width of the adder tree can become significantly wider. Consequently, achieving an optimal trade-off between performance, power, and area in conventional digital SRAM-CIM for high-precision PCN accelerations with real-time requirements proves challenging.

## III. Proposed PC2IM Architecture

To address memory access challenges in SP based PCNs and achieve high performance and energy efficiency in high-precision MLPs, this section introduces a novel PCN accelerator, named PC2IM, incorporated into a memory-efficient data flow facilitated by the customized CIM engines. Specifically, by leveraging approximate sampling and CIM-based data preprocessing module, PC2IM accelerates the data preprocessing stage in PCNs with significantly reduced off-chip and on-chip memory access. Additionally, an efficient CIM-based feature computing engine is proposed to attain high performance for high-precision MAC operations while maintaining high energy and area efficiency.

### *A. Overview of Proposed PC2IM*

The proposed PC2IM is shown in Fig. 3(a) and comprises data preprocessing and feature computing modules. The data preprocessing module is composed of a distance computing engine named approximate distance SRAM-CIM (APD-CIM), a two-level Ping-Pong-MAX content-addressable memory (Ping-Pong-MAX CAM) to store temporary minimal distance and find centroids, on-chip buffers for index and feature storage, and digital units includes sorter/merger and aggregation units, etc. The feature computing module consists of buffers, post processing units, and a feature computing engine named split-concatenate SRAM-CIM (SC-CIM) for efficient MLPs. PC2IM supports both point set abstraction (PSA) and point feature propagation (PFP) layers with delayed aggregation flow to reduce inter-layer feature sizes [8] as shown in Fig. 3(b).

### *B. Proposed CIM-based Data Preprocessing Module*

To address the ***Challenge I*** mentioned in Section II-B, a CIM-based data preprocessing module incorporated in a memory-efficient data flow is proposed to simultaneously alleviate both on-chip point access and temporary distance update overhead in point-based PCNs. The sampling flow in data preprocessing of PCNs is first optimized for CIM and further facilitated by the customized CIM engines with software-hardware co-design.

***Approximate sampling***. Distance between two three-dimensional points in PCNs is generally represented as Euler distance $L_2$ which is calculated as

$$R^2 = (x - x_r)^2 + (y - y_r)^2 + (z - z_r)^2 \qquad (1)$$

where ($x_r$, $y_r$, $z_r$) refers to the coordinates of a reference point and ($x$, $y$, $z$) refers to a certain query point. The main operators of $L_2$ are abstraction, multiplication, and addition, which can be power-hungry and inefficient for CIM. To this end, we propose a CIM-friendly approximate distance based sampling for spatial partitioned PC based on the median to efficiently conduct distance calculation in memory. $L_2$ in PCNs is approximate into Manhattan distance $L_1$ in this work which is calculated as

$$L = |x - x_r| + |y - y_r| + |z - z_r|. \qquad (2)$$

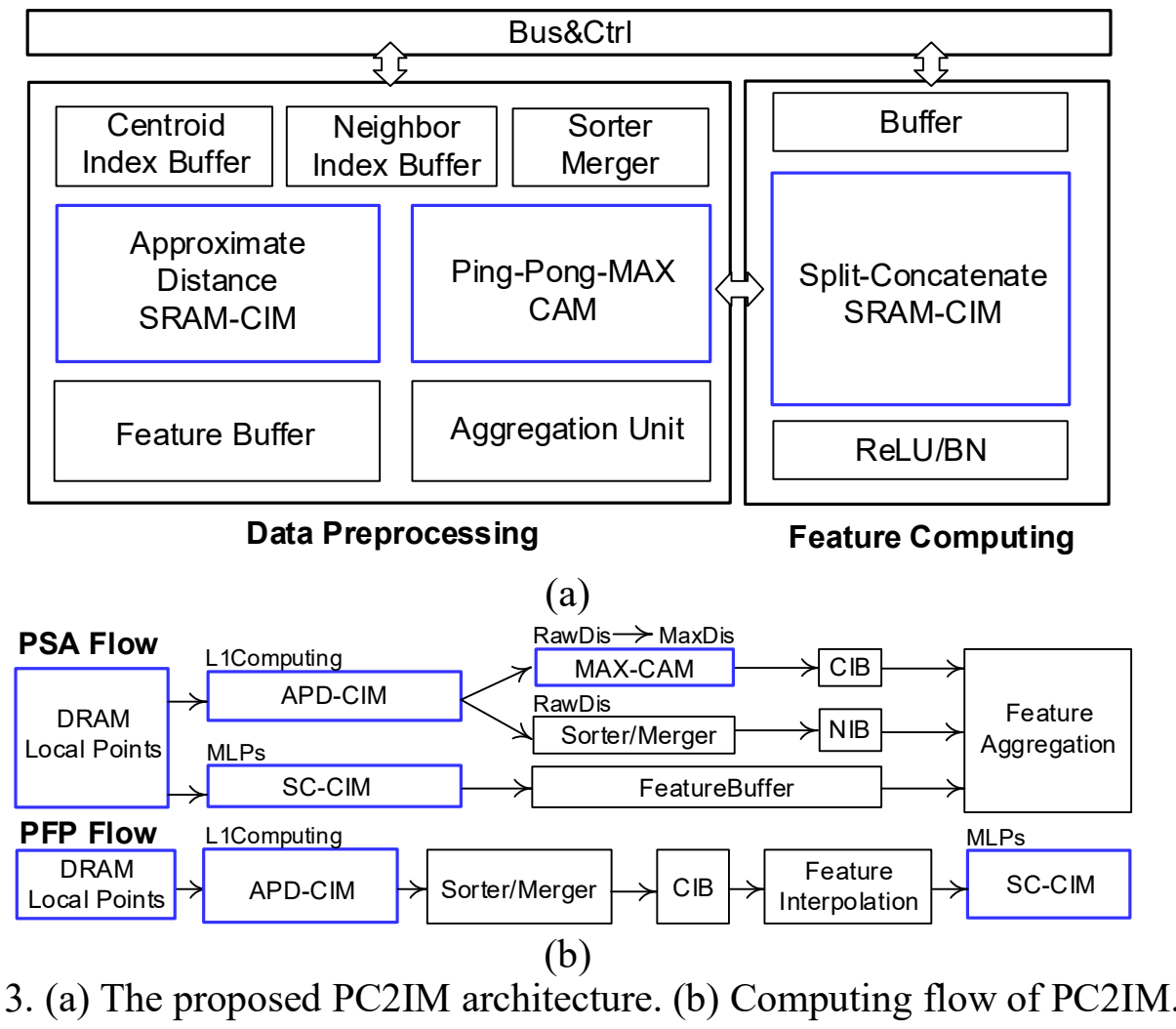

Fig. 3. (a) The proposed PC2IM architecture. (b) Computing flow of PC2IM.

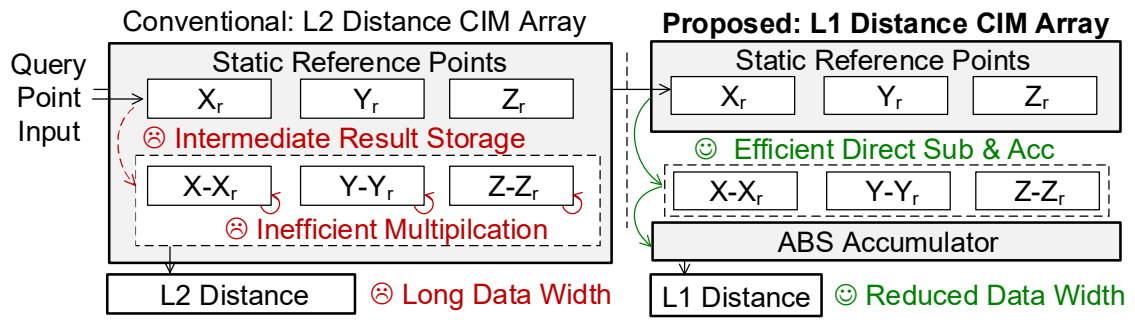

Fig. 4. Comparisons of computing flows between conventional $L_2$ distance CIM array and the proposed $L_1$ distance CIM array.

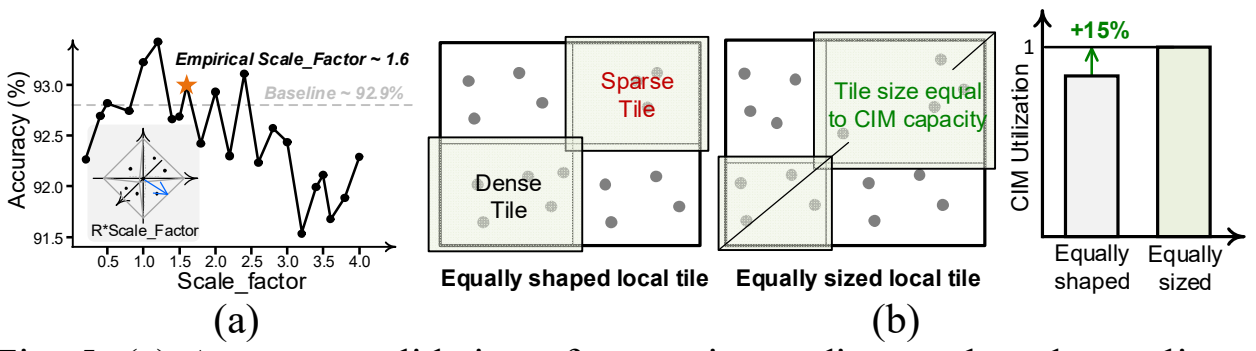

Fig. 5. (a) Accuracy validation of approximate distance based sampling. (b) Median based spatial partitioning with equally sized local tile to enhance on-chip CIM utilizations.

$L_1$ can be more efficient for CIM compared to $L_2$, as shown in Fig. 4. Detailly, $L_2$ calculation needs costly multipliers directly imbedded in or beside CIM arrays, leading to significant area and power overhead. Otherwise, the dynamic intermediate results (($x$-$x_r$), ($y$-$y_r$), and ($z$-$z_r$)) need to be used as operands for multi-cycle in-situ multiplication in array, which is inefficient for CIM that typically stores static data. More importantly, $L_2$ expands the data width by nearly 2×, increasing the hardware overhead of subsequent digital processing units and necessitating significantly larger data sizes for the temporary distance memory that needs updating. Note that, the approximation from $L_2$ to $L_1$ is feasible since they share the similar trending when the coordinates change in most of cases. Building upon $L_1$, the query range in neighbor search can be treated as lattice with an adaptive query range $L$, as shown in Fig. 5(a). $L$ is scaled by an empirical factor of 1.6 compared to the original ball query radius $R$ to ensure that there is no explicit information loss in neighbor search. This is validated in the small dataset ModelNet40, where the accuracy is well preserved compared to the previous Euler distance-based sampling method.

To enhance the hardware utilization of on-chip CIM array, the median-based spatial partitioning (MSP) is introduced, shown in Fig. 5(b). In MSP, local tiles are partitioned using the median along a certain axis, resulting in equally sized tiles with an unfixed shape. This enables local tiles to fit seamlessly into the on-chip CIM array, leading to an average 15% increased array utilization, as verified through evaluations on the S3DIS dataset. Besides, each equally sized tile samples the same number of centroids with uniformed access patterns for structured CIM hardware. MSP is executed by the host CPU initially and can be effectively accelerated using previously developed K-D tree accelerators [15].

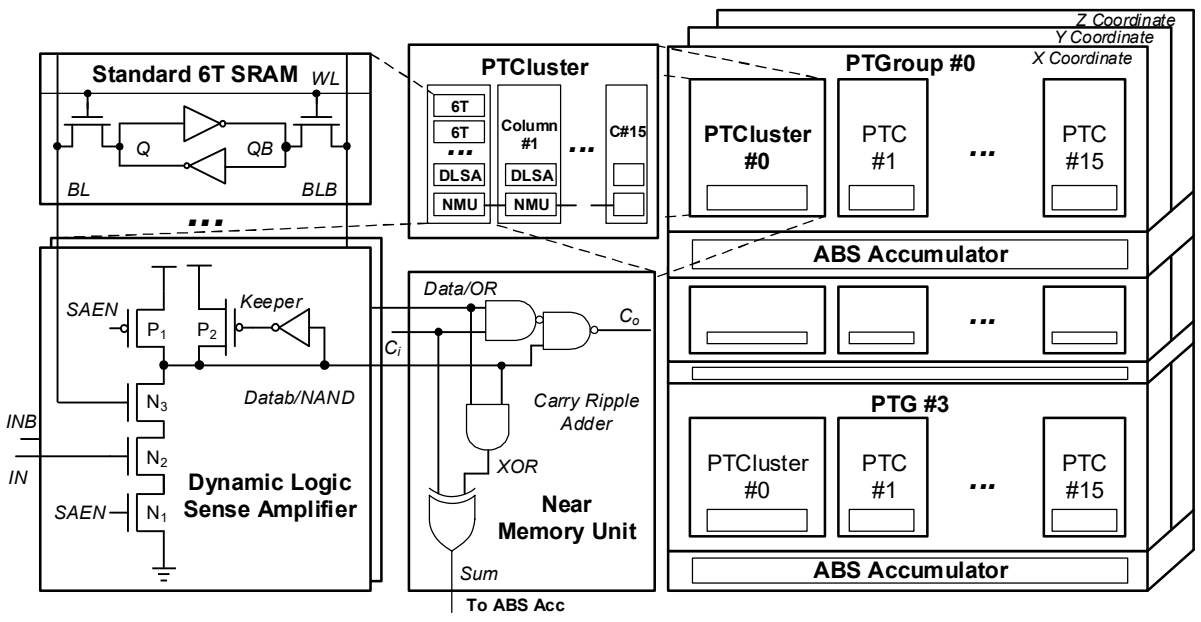

Fig. 6. The proposed APD-CIM.

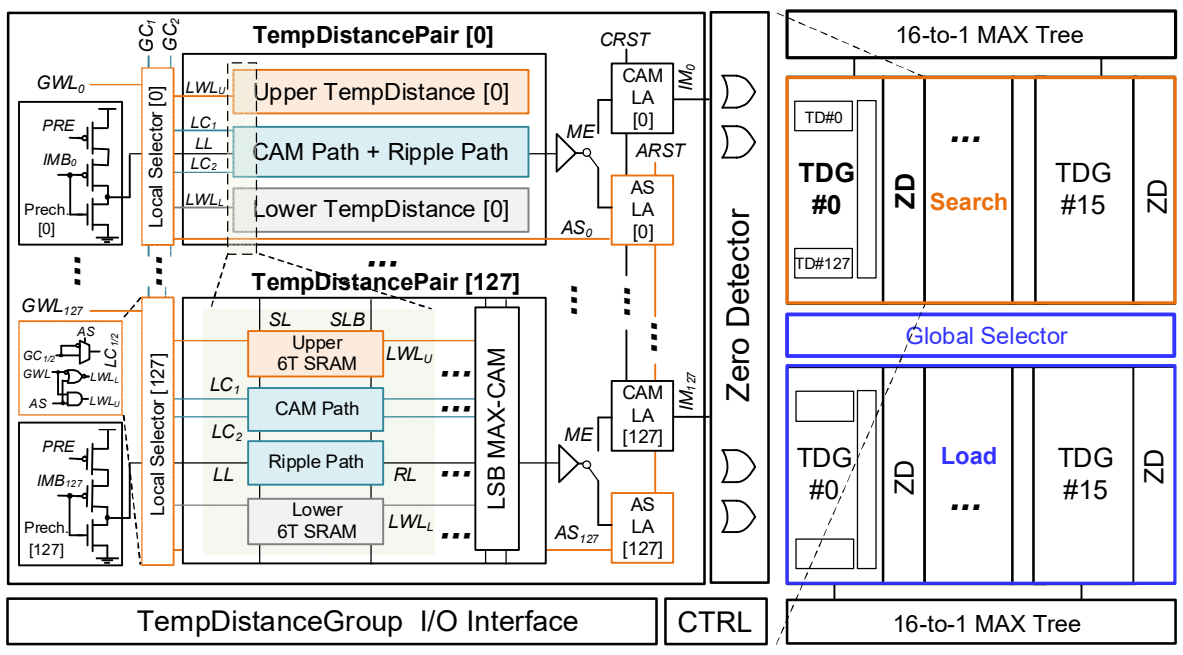

Fig. 7. The proposed two-level Ping-Pong-MAX CAM.

***Approximate Distance SRAM-CIM.*** The proposed approximate distance SRAM-CIM is used for $L_1$ distance calculations, as shown in Fig. 6. The APD-CIM array comprises four point groups (PTGs), each containing 16 point clusters (PTCs) with absolute value accumulators (ABSAcc). The PTC is composed of standard 6T SRAM for 32 points storage, dynamic logic sense amplifiers, and near memory units. Capable of storing up to 2048 points with 16-bit quantization, the APD-SRAM-CIM accounts for 12KB. The dynamic logic sense amplifier is used to conduct simple logic functions like NAND/OR. This done by adding additional pulldown transistor $N_2$ controlled by input signals. The near memory unit uses the NAND/OR results to implement additions. The abstraction is achieved by inverting inputs and setting $C_0$ to 1. The addition results are sent to ABS accumulators to output final results. To initiate distance computing, the reference point is firstly read out from PTGs and stored in registers for bit-parallel inputs. In each cycle, 16 19-bit $L_1$ distances are generated by activating one row of PTG and sent to MAX-CAM and sorter for FPS and lattice query, respectively.

***Ping-Pong-MAX CAM.*** The proposed two-level Ping-Pong-MAX CAM is used to find the maximal $\boldsymbol{D_s}$ to get the centroid indexes, as shown in Fig. 7. The Ping-Pong-MAX CAM is composed of a global selector and two CAM arrays with 16 temporary distance groups (TDGs) and 16-to-1 MAX tree for global maximal in each array. The global selector is used to selectively configure the CAM array into search mode or load mode for pipelined executions in an array-level Ping-Pong fashion. In each TDG, 128 pairs of temporary distances (TDPs) are stored in the paired MAX-CAM cells, which comprise of upper and lower SRAM cells controlled by local wordline $LWL_U$ and $LWL_L$ with shared CAM and ripple paths, as shown in Figs. 7 and 8(a). The ripple math line ($LL$ and $RL$) of MAX-CAM cell is connected to a sense amplifier and a precharger with a feedback control. The adaptive selector latch (AS-LA) and CAM result latch (CAM-LA) are used to latch up compare and CAM results, respectively. Besides, a zero detector composed of pure OR gates is used to detect match signals in each local TDG.

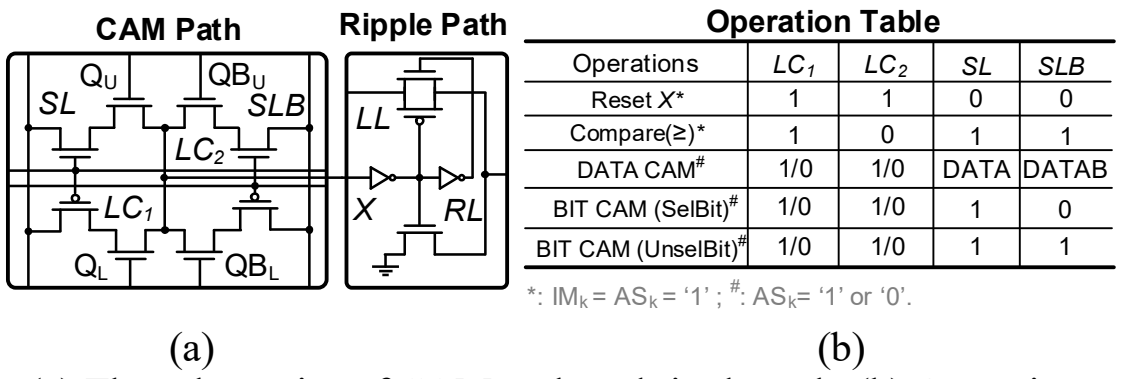

Fig. 8. (a) The schematics of CAM path and ripple path. (b) Operation table of proposed Ping-Pong-MAX CAM array.

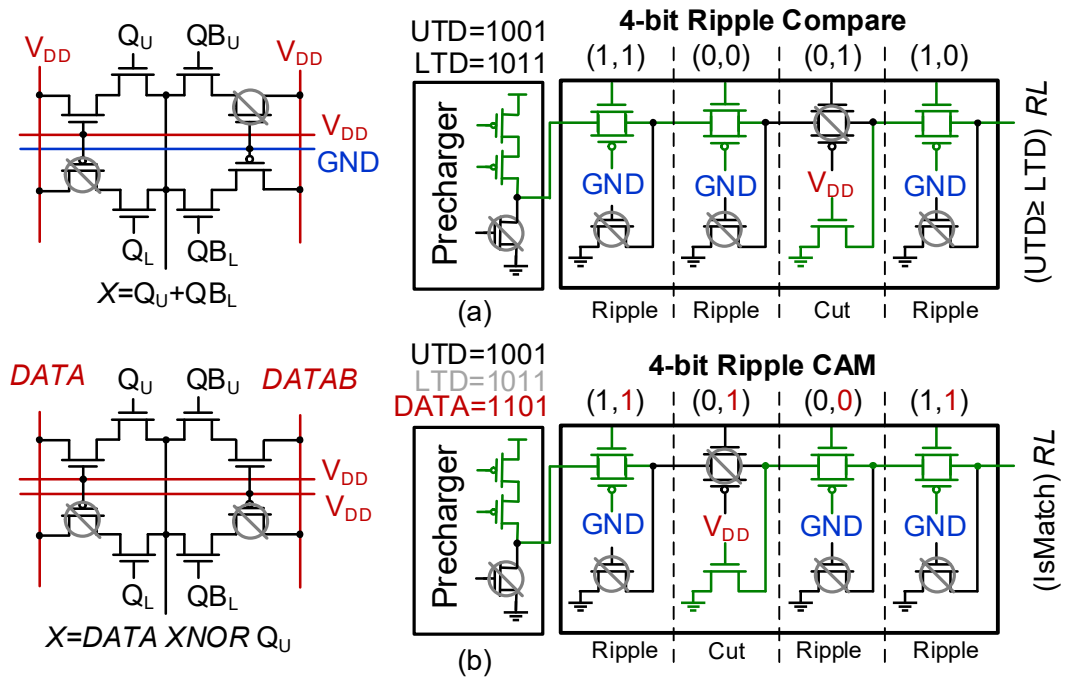

Fig. 9. (a) An example of 4-bit in-situ comparison for upper TD and lower TD. (b) An example of 4-bit data CAM with unmatched search patterns.

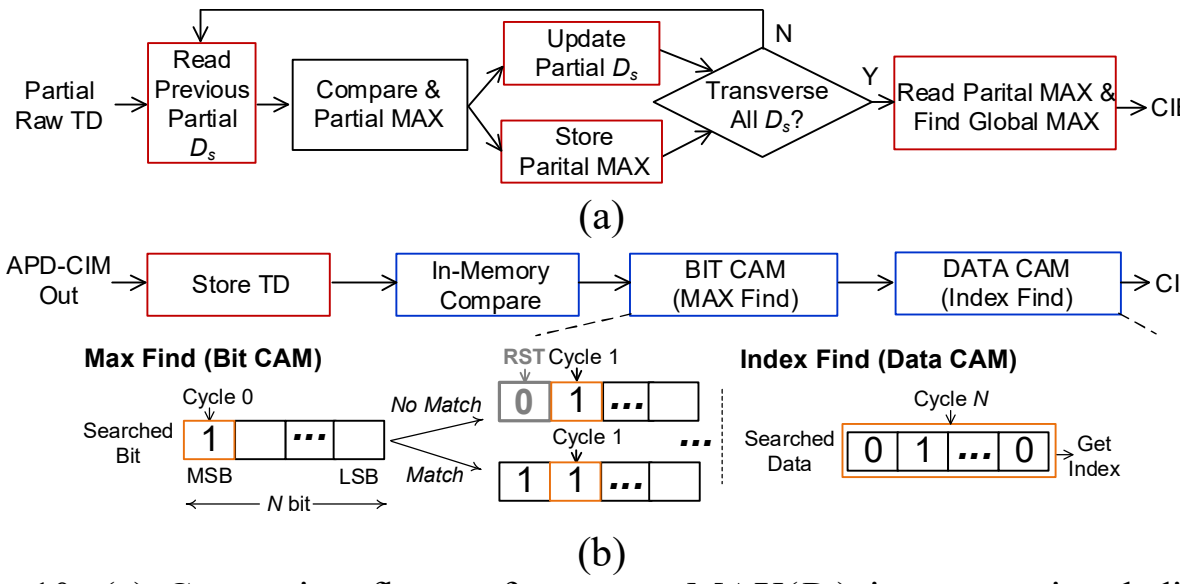

Fig. 10. (a) Computing flows of operator MAX($\boldsymbol{D_s}$) in conventional digital circuits with intensive memory access. (b) Computing flows of operator MAX($\boldsymbol{D_s}$) in proposed Ping-Pong-MAX CAM array.

The proposed MAX-CAM cell supports in-situ comparison and adaptive bit or data CAM operations between upper and lower TDs, as shown in Fig. 8(b). The in-situ comparison is finished by configuring the local configuration line $LC_1$ and $LC_2$ as '1' and '0', respectively, and ripple the first precharged $LL$ to the last $RL$, as shown in Fig. 9(a). The compare result is latched up in AS-LA. AS-LA then controls the local selector to convert the global signals $GWL$ and $GC_{1/2}$ to local signals $LWL_{U/L}$ and $LC_{1/2}$ that adaptively performs distance update on the larger TD while conducting bit or data CAM on the smaller TD in a cell-level Ping-Pong fashion.

The CAM array uses bit and data CAM modes to sample the desired centroid with maximal distance. The bit CAM is used to find the maximal 19-bit TD value stored in CAM array in 19 cycles from MSB to LSB. A mismatch in bit CAM will exclude current TDP from the following bit search cycles by disabling the precharger controlled by CAM-LA. After obtaining the maximal TD, the data CAM performs bit-parallel search to get the corresponding index, as shown in Fig. 9(b). Compared with conventional digital implementations, the proposed Ping-Pong-MAX CAM array alleviate the need of frequent memory access for TDs and partial MAX by performing in-situ comparison

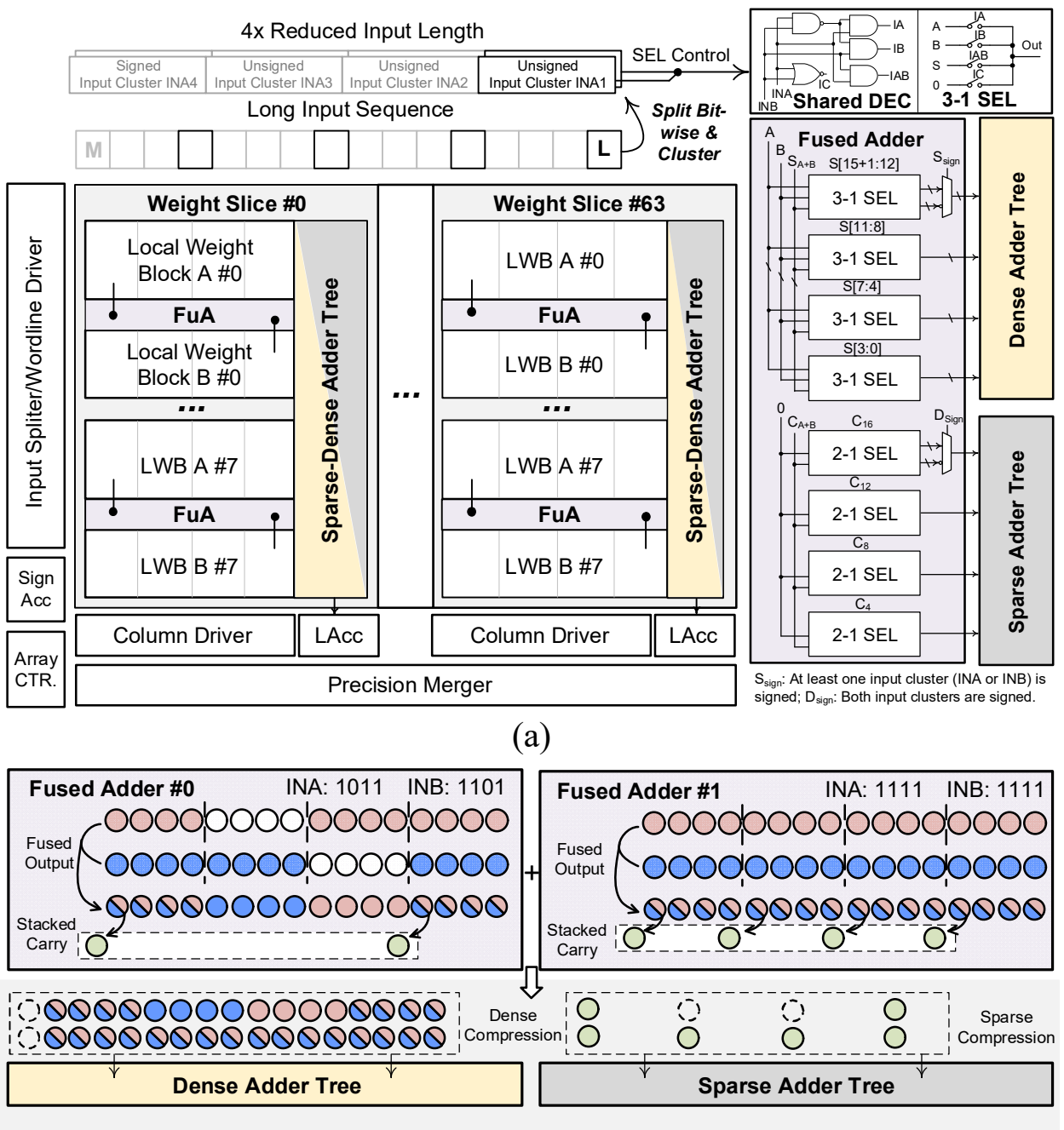

(a)

(b)

Fig. 11. (a) The proposed split-concatenate SRAM-CIM. (b) The fused multiply-and-add operations. Red circle: data from LWB A; Blue circle: data from LWB B; Fused circle: sum out from CRA; Green circle: carry from CRA.

and CAM in a memory-efficient data flow, as shown in Fig. 10.

### C. Proposed CIM-based Feature Computing Engine

To address the ***Challenge II*** mentioned in Section II-B, an efficient feature computing engine, named split-concatenate SRAM-CIM (SC-CIM), is proposed for high performance MAC operations while maintaining high energy and area efficiency, as shown in Fig. 11(a). SC-CIM is composed of 64 weight slices. In each weight slice, 8 paired 4-bit local weight blocks (LWB) share a dense-sparse adder tree for accumulating partial sums, where each weight block stores 16 rows of weights. The paired weight blocks A and B share a fused adder (FuA) for concatenate-based multiplication and additions with low area overhead. The FuA is essentially based on transmission gates and controlled by decoded inputs.

To initiate split-concatenate based MAC operations, the long input sequence is firstly split bit-wise in a 4-bit interleaved manner and grouped into 4-bit input clusters, while the weight is split block-wise in a 4-bit consecutive manner. In each cluster, the significance of adjacent input bit is increased by $2^4$ instead of $2^1$, as shown in the top of Fig. 11(a). For each computing cycle, the 4-bit input cluster is sent to CIM array with 4× reduced computing cycles compared to conventional bit-serial one. When the 4-bit input cluster multiplies 4-bit weight block (cluster-block multiplication), the primary operation is to select weight block to concatenate a 16-bit output without costly multipliers.

To reduce the accumulation overhead, we fuse the first accumulation stage with cluster-block multiplications, by sharing the concatenate units between adjacent weight blocks A and B to form the FuA, which is controlled by two adjacent input clusters INA and INB. FuA is composed of a 4-bit carry ripple adder (CRA, not shown in Fig. 11(a)), 3-1 and 2-1 SEL units. CRA is used to add the 4-bit weight from LWBs A and B in advance, regardless of input patterns. The 3-1 SEL with an extra sign extension bit is used to form a densely concatenated (16+1)-bit output by selecting the 4-bit weight from LWBs or the 4-bit sum from CRA according to the decoded signed/unsigned input clusters from the array shared decoder. The dense output is then accumulated by dense adder tree. The 1-bit carry signal from CRA is sparsely concatenated by 2-1 SEL and accumulated within sparse adder tree. To address the sign extension problem of concatenating a signed weight block, we separately concatenate the signed and unsigned parts of weight blocks and merge them in the periphery. The fused multiply-and-add operation is shown in Fig. 11(b). The fused operation reduces accumulating number by half, resulting in nearly 44% reduced hardware overhead compared to the naive implementation of directly accumulating the large wide-width partial sums.

## IV. Evaluation Results

### A. Experimental Setup

***Algorithm.*** To validate the approximate sampling method with lattice query and MSP, three datasets and a representative point-based PCN named PointNet2 are selected for evaluations, as listed in Table I. The datasets cover tasks ranging from classification to segmentation with varied number of points, ensuring generality.

***Hardware.*** We implement the digital units in PC2IM using Verilog and synthesize it using Synopsys Design Compiler under 40nm technology with a clock frequency of 250MHz. The in-memory computing circuits are customized in Cadence Virtuoso with post-layout simulations. The detailed hardware specifications are provided in Table II. The on-chip point capacity is assumed to be 2k with 16-bit quantization.

***Baseline.*** We implement two baselines under 40nm technology, along with an additional commercial GPU baseline:

- Baseline-1: digital units with global PC access for data preprocessing + near memory computing for MLPs.
- Baseline-2 (DAC'23 [10]): digital circuits with local PC access for data preprocessing+ near memory computing for MLPs.
- Baseline-3 (General computing platform): a server GPU (RTX 4090) tested with build-in tools.

Additionally, we adopt two baselines with different storage-compute ratios (SCR) to verify the effectiveness of SC-SRAM-CIM, where SCR is defined as the number of rows of SRAM sharing a digital multiplier unit:

- BS-CIM (conventional): bit-serial digital SRAM-CIM processing 1-bit input each cycle.
- BT-CIM (ISSCC'22 [14]): booth coding based digital SRAM-CIM.

### B. Evaluation Results

***Accuracy.*** The accuracies of PointNet2 on various datasets under the proposed approximate sampling method with post-training 16-bit quantization (PTQ) are shown in Fig. 12(a). The suggested approximate distance method, coupled with MSP, results in a negligible impact on accuracy in large dataset, with less than a 2% loss. Additionally, the 16-bit quantization error is minimal, causing less than a 0.3% accuracy degradation.

***Data preprocessing module.*** The energy consumptions of various data preprocessing modules are assessed and compared in Fig. 12(b). The energy consumptions are normalized to baseline-1. The proposed PC2IM exhibits the most energy-efficient design, achieving a reduction of up to 97.9% and 73.4% in energy consumption for large-scale PCs compared to baseline-1 and -2, respectively. This efficiency is credited to

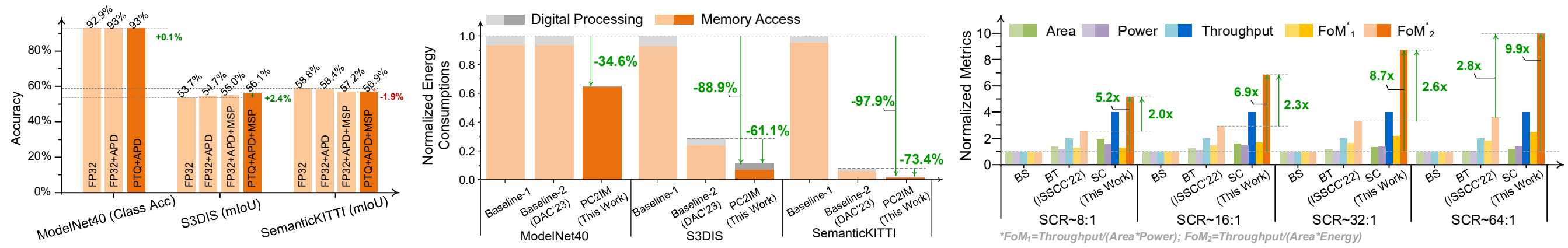

(a) (b) (c)

Fig. 12. (a) Software validation of accuracy results. (b) Energy consumption of data preprocessing. (c) Design metrics of various digital SRAM-CIM for feature computing under different storage-compute ratios (SCRs).

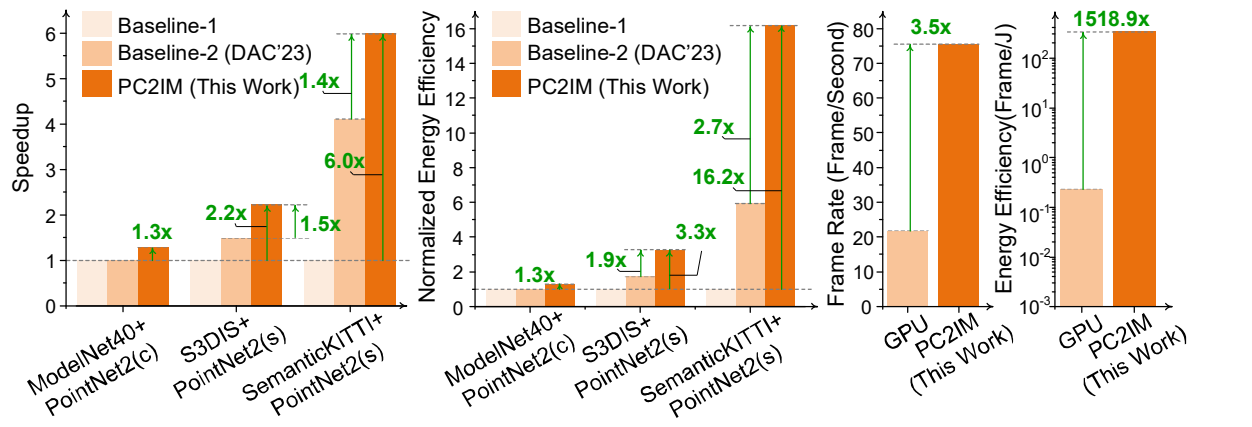

(a) (b) (c)

Fig. 13. System-level evaluations of (a) performance and (b) energy efficiency. (c) Comparisons with GPU on SemanticKITTI.

TABLE I
MODELS AND DATASETS

| Task | Dataset | # Points | PC model |
|---|---|---|---|
| Classification | ModelNet | 1k (small) | PointNet2 (c) |
| Semantic Segmentation | S3DIS | 4k (medium) | PointNet2 (s) |
| Semantic Segmentation | SemanticKITTI | 16k (large) | PointNet2 (s) |

TABLE II
HARDWARE SPECIFICATIONS OF PC2IM

| | | |
|---|---|---|
| **Technology** | | 40 nm |
| **Frequency** | | 250 MHz |
| **In-Memory Computing** | APD-CIM | 12 KB |
| | Ping-Pong-MAX CAM | 19 KB |
| | SC-CIM | 256 KB |
| **Standard On-chip SRAM** | | 512 KB |
| **Memory Access Energy*** | On-chip SRAM | 0.7 pJ/bit |
| | Off-chip DRAM | 4.5 pJ/bit |
| **Throughput (TOPS)** | | 2 (16b) |
| **Energy Efficiency (TOPS/W)** | | 2.53 (16b) |

*Characterized by CACTI 6.0 [16], SRAM-DRAM ratio is within the range of [13].

the CIM-based data preprocessing module incorporated into a memory-efficient data flow, significantly reducing both off-chip and on-chip memory accesses.

***Featuring computing module.*** The normalized design metrics of various digital SRAM-CIMs under different SCRs for feature computing are presented in Fig. 12(c). Thanks to the split-concatenate method, the proposed SC-CIM significantly boosts the throughput of conventional bit-serial manner by 4× with saved computing energy, resulting in 5.2× and 2.0× enhanced Figure of Merit 2 ($FoM_2$) at low SCR equal to 8, compared to BS-CIM and BT-CIM, respectively. With increased SCR, the area overhead of computing units is amortized by the memory array. This amplifies the benefits of SC-CIM, leading to up to 9.9× and 2.8× enhanced $FoM_2$ compared to BS-CIM and BT-CIM, respectively.

***Architecture-level analysis.*** The architecture-level results are shown in Fig. 13. Compared with previous PCN accelerator [10], the proposed PC2IM achieves 2.7× enhanced energy efficiency in large-scale dataset, attributed to the implementations of CIM-based data preprocessing module (contributing to 48.5%) and feature computing engine (contributing to 51.5%). With reduced off-chip memory access and efficient MLPs, PC2IM achieves speedups of up to 1.5× and 6.0× compared to baseline-1 and -2 designs, respectively, thanks to the integration of MSP and SC-CIM. Furthermore, in comparison with GPU implementation, PC2IM achieves a 3.5× speedup and 1518.9× enhanced energy efficiency, showcasing the efficacy of its efficient data preprocessing and feature computing scheme.

## V. CONCLUSION

In this paper, we introduce an efficient SRAM accelerator named PC2IM to alleviate off-chip and on-chip memory access of point-based 3D point cloud networks with enhanced performance. The proposed CIM-based data preprocessing module implementing a memory-efficient data flow reduces the energy consumptions of data preprocessing in PCNs by up to 73.4% compared to state-of-the-art methods. Additionally, the split-concatenate SRAM-CIM based feature computing engine is proposed for efficient high-precision MACs, achieving 2.0× to 9.9× enhanced FoM compared to previous digital SRAM-CIM. Architectural-level analysis demonstrates that the proposed PC2IM achieves 1.5× speedups and 2.7× enhanced energy efficiency compared to state-of-the-art PCN accelerator. Moreover, PC2IM achieves 3.5× and 1518.9× enhanced performance and energy efficiency compared to GPUs.

## REFERENCES

[1] Charles R. Qi *et al.*, "Pointnet++: Deep hierarchical feature learning on point sets in a metric space." *Advances in Neural Information Processing Systems (NeurIPS)*, Vol. 30, 2017.

[2] Y. Zhou and O. Tuzel, "Voxelnet: End-to-end learning for point cloud based 3d object detection," *Proceedings of the IEEE Conference on Computer Vision and Pattern Recognition (CVPR)*, June 2018.

[3] M. Han *et al.*, "QuickFPS: Architecture and Algorithm Co-Design for Farthest Point Sampling in Large-Scale Point Clouds," *IEEE Transactions on Computer-Aided Design of Integrated Circuits and Systems*, Vol. 42, No. 11, pp. 4011-4024, Nov. 2023.

[4] Y. Lin *et al.*, "PointAcc: Efficient point cloud accelerator," *Annual IEEE/ACM International Symposium on Microarchitecture (MICRO)*, pp. 449–461, October 2021.

[5] X. Yang *et al.*, "An Efficient Accelerator for Point-based and Voxel-based Point Cloud Neural Networks," *ACM/IEEE Design Automation Conference*, pp. 1-6, July 2023.

[6]S. Kim *et al.*, "PNNPU: A 11.9 TOPS/W High-speed 3D Point Cloud-based Neural Network Processor with Block-based Point Processing for Regular DRAM Access," *Symposium on VLSI Circuits*, pp. 1-2, June 2021.

[7]Z. Song *et al.*, "PRADA: Point Cloud Recognition Acceleration via Dynamic Approximation," *Design, Automation & Test in Europe Conference & Exhibition (DATE)*, pp. 1-6, April 2023.

[8]Y. Feng *et al.*, "Mesorasi: Architecture Support for Point Cloud Analytics via Delayed-Aggregation," *Annual IEEE/ACM International Symposium on Microarchitecture (MICRO)*, pp. 1037-1050, October 2020.

[9] X. Liu *et al.*, "FusionArch: A Fusion-Based Accelerator for Point-Based Point Cloud Neural Networks," *Design, Automation & Test in Europe Conference & Exhibition (DATE)*, pp. 1-6, April 2024.

[10]J. Zheng *et al.*, "TiPU: A Spatial-Locality-Aware Near-Memory Tile Processing Unit for 3D Point Cloud Neural Network," *ACM/IEEE Design Automation Conference (DAC)*, pp. 1-6, July 2023.

[11] X. Liu *et al.*, "MoC: A Morton-Code-Based Fine-Grained Quantization for Accelerating Point Cloud Neural Networks," *ACM/IEEE Design Automation Conference (DAC)*, pp. 1-6, July 2024.

[12] Hyunsung Yoon et al., "Fused Sampling and Grouping with Search Space Reduction for Efficient Point Cloud Acceleration," *ACM/IEEE Design Automation Conference (DAC)*, pp. 1-6, July 2024.

[13]F. Yu *et al.* "Crescent: taming memory irregularities for accelerating deep point cloud analytics," *Proceedings of the 49th Annual International Symposium on Computer Architecture (ISCA)*, pp. 962-977, June 2022.

[14]F. Tu *et al.*, "A 28nm 29.2TFLOPS/W BF16 and 36.5TOPS/W INT8 Reconfigurable Digital CIM processor with unified FP/INT pipeline and bitwise in-memory booth multiplication for cloud deep learning acceleration," *ISSCC*, pp. 1-3, February 2022.

[15]R. Pinkham *et al.*, "QuickNN: Memory and Performance Optimization of k-d Tree Based Nearest Neighbor Search for 3D Point Clouds," *HPCA*, pp. 180-192, February 2020.

[16]N. Muralimanohar, R. Balasubramonian, and N. Jouppi, "Cacti 6.0: A tool to model large caches," HP Laboratories, 01 2009.